# Scalar and Tensor Sea Contributions to Nucleons

M. Batra[1*] and A. Upadhayay[2]
*SPMS, Thapar University Patiala-147004.*
*email:mbatra310@gmail.com

## Introduction:

The low energy parameters of baryons like spin distribution, magnetic moment, weak decay, coupling constant ratios are still a challenge for particle physicist. In past various models and theories have been suggested to describe the structure of baryons in these context. The experimental investigation for the structure of baryons came into existence from Deep-Inelastic experiments. Later on EMC [1] at CERN concluded that spin of proton is carried by quark -anti quark pair in addition to valence part. SMC collaboration [2] measured the spin structure $g_1(x)$ of proton and neutron at $Q^2=10$ GeV$^2$. Also different models are put forward with perturbative and non perturbative approach to define the quark interactions. For example, Isgur and Karl [3] defined the confining interaction potentials among quarks. Some models explain the low energy properties through presence of constituent quarks inside the baryon. In order to have reliable information about the various properties of hadrons at low energy, self-consistent approach should be adopted. Low energy parameters for baryons are computed here by assuming baryons as a composite system of three quarks known as valence part and sea-part as made up of quark-antiquark pair and gluons. The wave-function [4] for valence part of the baryon is denoted by $\Psi = \Phi[|\phi\rangle|\chi\rangle|\psi\rangle]|\xi\rangle$ where each part contributes in such a way so as to make the total wave-function anti-symmetric in nature. Here sea is assumed to be flavorless and in S-wave, spin of sea part are chosen to be in 0, 1, 2 and color can be $(1_c, 8_c, \overline{10}_c)$. The total flavor-spin-color wave function of a spin up baryon which consist of three-valence quarks and sea components can be written as

$$\left|\Phi_{\frac{1}{2}}^{\uparrow}\right\rangle = \frac{1}{N}[\Phi_1^{(\frac{1}{2})\uparrow} H_0 G_1 + a_8 \Phi_8^{(\frac{1}{2})\uparrow} H_0 G_8 + a_{10}\Phi_{10}^{(\frac{1}{2})\uparrow} H_0 G_{\overline{10}} + b_1 \left[\Phi_1^{\frac{1}{2}}\otimes H_1\right]^{\uparrow} G_1 +$$
$$+ b_8 \left(\Phi_8^{\frac{1}{2}}\otimes H_1\right)^{\uparrow} G_8 + b_{10}\left(\Phi_{10}^{\frac{1}{2}}\otimes H_1\right)^{\uparrow} G_{\overline{10}} + c_8(\Phi_8^{\frac{3}{2}}\otimes H_1)^{\uparrow}G_8 + d_8(\Phi_8^{\frac{3}{2}}\otimes H_2)^{\uparrow}G_8]$$

Where $\quad N^2 = 1 + a_8^2 + a_{10}^2 + b_1^2 + b_8^2 + b_{10}^2 + c_8^2 + d_8^2$

Each term in above defined wave-function consists of two parts where $\Phi$ represents contribution from valence part and other contributes to sea. Various combinations contributing to sea take into account antisymmetry of each term in the wave-function. The first three terms in wave-function come from coupling of spin 1/2 state coupled to scalar sea and other three $b_1$, $b_8$, $b_{10}$ represents coupling between spin 1/2 and vector Sea. Finally terms with $c_8$ and $d_8$ are due to coupling of spin 3/2 with vector and tensor sea respectively. Various low energy parameters can be calculated by defining a suitable operator for each property of the system.

$$\left\langle\Phi_{1/2}^{(\uparrow)}\middle|\hat{O}\middle|\Phi_{1/2}^{(\uparrow)}\right\rangle = \frac{1}{N^2}[a\sum_i <\hat{O}_f^i>^{\lambda\lambda}<\hat{\sigma}_z^i>^{\lambda\uparrow\lambda\uparrow} + <\hat{O}_f^i>^{\rho\rho}<\hat{\sigma}_z^i>^{\rho\uparrow\rho\uparrow} + 2 <\hat{O}_f^i>^{\lambda\rho}<\hat{\sigma}_z^i>^{\lambda\uparrow\rho\uparrow} \quad +b \sum_i(<\hat{O}_f^i>^{\lambda\lambda} + <\hat{O}_f^i>^{\rho\rho})(<\hat{\sigma}_z^i>^{\lambda\uparrow\lambda\uparrow}+<\hat{\sigma}_z^i>^{\rho\uparrow\rho\uparrow}) + c\sum_i[<\hat{O}_f^i>^{\lambda\lambda}<\hat{\sigma}_z^i>^{\rho\uparrow\rho\uparrow} + <\hat{O}_f^i>^{\rho\rho}<\hat{\sigma}_z^i>^{\lambda\uparrow\lambda\uparrow} - 2<\hat{O}_f^i>^{\lambda\rho}<\hat{\sigma}_z^i>^{\rho\uparrow\rho\uparrow}] + d[\sum_i<\hat{O}_f^i>^{\lambda\lambda}+\sum_i<\hat{O}_f^i>^{\rho\rho}] + e[\sum_i(<\hat{O}_f^i>^{\rho\rho}-<\hat{O}_f^i>^{\lambda\lambda})<\hat{\sigma}_z^i>^{\lambda\uparrow\frac{3}{2}\uparrow} + 2\sum_i<\hat{O}_f^i>^{\lambda\rho}<\hat{\sigma}_z^i>^{\rho\uparrow\frac{3}{2}\uparrow}]$$

Here $<\hat{O}_f^i>^{\lambda\lambda}=\langle\phi^\lambda|O_f^i|\phi^\lambda\rangle$, $<\hat{\sigma}_z^i>^{\lambda\uparrow\lambda\uparrow}= \langle\chi^{\lambda\uparrow}|\sigma_z^i|\chi^{\lambda\uparrow}\rangle$, $<\hat{O}_f^i>^{\rho\rho} = \langle\phi^\rho|O_f^i|\phi^\rho\rangle$ and $<\hat{\sigma}_z^i>^{\rho\uparrow\rho\uparrow}= \langle\chi^{\rho\uparrow}|\sigma_z^i|\chi^{\rho\uparrow}\rangle$

$$\langle\chi^{\lambda\uparrow}|\sigma_z^i|\chi^{\lambda\uparrow}\rangle = \left(\frac{1}{\sqrt{6}}(\uparrow\downarrow+\downarrow\uparrow)\uparrow-2\uparrow\uparrow\downarrow\middle|\sigma_z^i\middle|\frac{1}{\sqrt{6}}(\uparrow\downarrow+\downarrow\uparrow)\uparrow-2\uparrow\uparrow\downarrow\right) = \frac{2}{3} \qquad (1.1)$$

$\widehat{O}_f^i$ depends upon the flavor of $i^{th}$ quark and $\hat{\sigma}_z^i$ spin projection in z-direction operator of $i^{th}$ quark. The suitable operators are switched for flavor and spin so as to get a simplified relation for each of the parameter in the wave-function, for instance a spin projection operator(eqn.1.1) when operates on symmetric part of proton wave function gives a value 2/3. Low energy properties for proton and neutron are discussed here by using two approaches, one is statistical model proposed by Singh and Upadhyay [5] while other approach, firstly proposed by Li[6] and later on extended by Song and Gupta[4]. They computed low energy properties of nucleons by evaluating $\alpha$ and $\beta$ from the experimental data of magnetic moment with suppressed scalar & tensor sea. Singh and Upadhyay [5]considered proton as an ensemble of quark-gluon states in terms of sub processes q⇔qg, g⇔gg and g⇔q$\bar{q}$ and calculated the probability associated with different Fock states coming from scalar, vector and tensor sea. The relative probability in spin and color space is taken for each of these case so as the core part to have an angular momentum as $j_1$ and sea-content to have final angular momentum as $j_2$ and the total angular momentum should be $j_1 + j_2 = 1/2$ and resultant should be a color singlet state. The symmetry conditions wherever necessary are also applied.. Using these probability ratios, different coefficients are calculated where all coefficients are expressed in relations in terms of $\alpha$ and $\beta$ [4].

**Results and Conclusion:**

The method stated above use a non-relativistic approach for proton and neutron system assuming nucleons as a composite system of three quarks and "sea" where sea is assumed to be in S-wave consisting of quark-antiquark pairs ($u\bar{u}, d\bar{d}$ only) and gluons. Here the contributions from scalar, vector and tensor sea is taken into consideration to find $\alpha$ and $\beta$ [4]. Distribution of spin among valence quarks and sea-quark for proton and neutron are denoted by $I_1^p$ and $I_1^n$ respectively. Statistical model predicted the values as 0.132 and -0.011 which is close to EMC experimental value 0.136 and -0.013[7] for $I_1^p$ and $I_1^n$ respectively. Although the statistical model predictions give close to EMC data for neutron spin distribution but Song and Gupta results matches with SU(6) model results. Similarly for magnetic moment ratio of proton to neutron, the predicted value is -1.40 by statistical model which again matches with the PDG value [8] i.e. -1.459. The weak decay coupling ratio can be measured using F and D, for $(g_A/g_V) n \rightarrow p$ = F+D where F and D are weak matrix elements in beta-decay. Using Bjorken sum rule, weak decay coupling constant is found to be equal to 1.50($\alpha_s$ =0.5) in statistical model whereas using Song and Gupta approach; we have the value 1.25 therefore both the values lie within the range ~2-20% of the experimental value[8] 1.26±0.00028. In summary, the results from two methods lie close to the allowed range for each low energy parameter. The Statistical model approach is favorable since it consider the contributions from scalar, vector and tensor sea unlike Song and Gupta approach, which suppresses the contribution from scalar and tensor sea. Hence we point the need for inclusion of most of the sea contribution to make the analysis in better agreement with the existing experimental data.